# Birefringence effect in Raman scattering of anisotropic materials

Muqian wen

## Abstract

This paper adds birefringence consideration to the conventional classical Raman scattering theory commonly found in literature to address birefringence and interference effects in Raman spectroscopy. It is hoped that the new results can be useful for ab initio calculations as well as experimental studies.

## Introduction

Birefringence effects in Raman spectroscopy is a useful phenomenon. For example, it can be used to study stress in crystals [1,2]. Many ab initio studies of Raman spectra often cite the two reference books by P. Brüesch and M. Cardona for the equations to calculate Raman intensities [3,4]. However, the cited equations could not account for birefringence and interference effects, which arise from anisotropy and finite thicknesses of the samples, nor can it be apparent from the derivations on how to account for them. These equations were derived before layered materials had become a thing. Layered materials are often anisotropic and few-layer thin, where birefringence and interference can have severe impacts. It is well known that ab initio calculated Raman spectra will deviate from experimental result to some extent for various reasons [5-26]. Birefringence and interference are one of the them. While there have been efforts to add birefringence correction to angle resolved polarized Raman spectroscopy studies of 2D materials and anisotropic crystals in the literature [5-10], those efforts are of empirical nature with some parameters fitted from experiment results and thus is not ideal for ab initio calculations. This paper would redevelop the classical Raman scattering theory from the ground up by following similar derivation procedures as the references [3,4] to address these shortcomings. Compared to the literature the derivations of this paper can be better suited for ab initio calculations and offer clearer physical pictures. And crucially the conclusions differ with the key difference being that the literature would approximate Raman active phonon wave vectors to 0 in derivation while this paper would not.

## Theory

When light propagates through a birefringent material, it will split into two mutually orthogonal polarized parts, each with different propagation speed. One is ordinary ray and the other is extraordinary ray. In biaxial birefringent materials both are extraordinary rays. But in this paper, we can still denote them as o ray and e ray for convenience. According to classical electrodynamics, the electric field of an incident light $\vec{E}$ will induce a dipole moment $\vec{\mathrm{M}}$ in an atom.

$$\vec{\mathrm{M}} = \varepsilon_0 \overleftrightarrow{x} \vec{E} \qquad (1)$$

Where $\varepsilon_0$ is vacuum permittivity. $\overleftrightarrow{x}$ is polarizability tensor.

The behavior of light waves in anisotropic materials is relatively more complex [11]. In birefringent materials the electric displacement field is generally not in the same direction as the direction of the electric field. The electric displacement field is the product of the dielectric tensor and the electric field vector. In birefringent materials the electric displacement field will be perpendicular to the propagation direction, but the electric field will not. The electric field is perpendicular to the energy flow direction which is usually not the same direction as the wave propagation direction in birefringent materials. Furthermore, in birefringent materials, an incident polarized light wave from outside will in general split into two refracted waves propagating at different speeds. And for each given propagation direction, there will be two possible propagation speeds corresponding to two mutually perpendicular electric displacement field directions. This property is all one needs to know about birefringent light propagations in this paper.

Assuming that the incident light is a simple monochromatic plane wave, then for the k-th atom in the f-th unit cell of the crystal, the dipole moment becomes

$$\vec{M}_{fk} \propto \overleftrightarrow{x}_{fk} A_a \hat{e}_a e^{i(\omega_i t - n_a \vec{k}_a \cdot \vec{r}_{fk})} \qquad (2)$$

Where $\hat{e}_a$ is the unit vector of the electric field direction $a$, which must be in one of the electric field directions of o ray or e ray light waves in birefringent materials, and $n_a$ is the corresponding refractive index, $A_a$ is amplitude, $\vec{k}_a$ is vacuum wavenumber. $\omega_i$ is the frequency of incident light which would be independent of polarization direction. $\vec{r}_{fk}$ is the coordinate of $fk$-th atom. The electric field in the above equation is expressed in the complex form and likewise the resulting dipole moment in the above equation will also be complex. This is certainly not the reality, and we should only take the real part of it for the actual dipole moment. The dipole moment equation must be expressed in complex form because the real sinusoidal functions forms will not be able to express phase additions easily. Both the polarizability tensor and the amplitude can have additional phases and thus will be complex in general.

The polarizability tensor can be written in two parts:

$$\overleftrightarrow{x}_{fk} = \overleftrightarrow{x}_k^0 + \overleftrightarrow{x}_{fk}^p \qquad (3)$$

The first part $\overleftrightarrow{x}_k^0$ is independent of time and position of unit cell. It is responsible for Rayleigh scattering. The second part is the polarizability induced by atom movement

$$\overleftrightarrow{x}_{fk}^p = \frac{\partial \overleftrightarrow{x}_k}{\partial \vec{\mathrm{r}}_k} \cdot \vec{u}_{fk} \propto \frac{\partial \overleftrightarrow{x}_k}{\partial \vec{\mathrm{r}}_k} \cdot \left( \frac{\vec{g}_k(\vec{q},j)}{\sqrt{m_k}} A_{(\vec{q},j)} e^{i(\omega(\vec{q},j)t - \vec{q} \cdot \vec{r}_{fk})} + c.c. \right) \qquad (4)$$

where $\vec{u}_{fk}$ is displacement of atom. $\vec{g}(\vec{q},j)$ is the eigen vector of a particular vibration mode with wave vector $\vec{q}$ and frequency $\omega(\vec{q},j)$. $m_k$ is the atomic mass. $A_{(\vec{q},j)}$ is complex amplitude of the lattice vibration mode. This part is responsible for Raman scattering. Now, according to classical radiation theory an oscillating dipole with frequency $\omega$ will emit energy per unit solid angle at a rate of [3,4]:

$$\frac{\mathrm{dW}}{\mathrm{d}\Omega}=\frac{\omega^4}{16\pi^2\varepsilon_0 c^3}\left|\hat{e}\cdot\vec{M}\right|^2 \qquad (5)$$

Hence, each atom will emit a scattered light of:

$$\vec{E}_{fk}^{b}\propto\vec{M}_{fk}^{b}e^{in_b\vec{k}_b\cdot\vec{r}_{fk}} \qquad (6)$$

where $b$ denotes the polarization or oscillation direction of the scattered light, which also must be in either one of the o ray or e ray electric field directions of the propagation direction of the specific Raman scattering.

The total Raman scattered light by a particular vibration mode will be all these individual scattered lights added up:

$$\vec{E}_s\propto\sum_{fkabd}\hat{e}_b\left(\hat{e}_b\cdot\frac{\partial\vec{\overleftrightarrow{x}}_k}{\partial r_{kd}}\cdot\hat{e}_a\right)A_a e^{i(\omega_i t+(n_b\vec{k}_b-n_a\vec{k}_a)\cdot\vec{r}_{fk})}\left(\frac{g_{kd}(\vec{q},j)}{\sqrt{m_k}}A_{(\vec{q},j)}e^{i(\omega(\vec{q},j)t-\vec{q}\cdot\vec{r}_{fk})}+c.c.\right) \qquad (7)$$

where the annotation $d$ denotes the axis direction which can be any one of the x, y, z. $\hat{e}_b$ is unit vector at $b$ polarization direction. It can be seen from the above equation that for bulk crystals $\vec{q}$ and $\omega(\vec{q},j)$ should satisfy:

$$\begin{gathered}\pm\vec{q}=Re(n_a)\vec{k}_a-Re(n_b)\vec{k}_b\\ \pm\omega(\vec{q},j)=\omega_i-\omega_s\end{gathered} \qquad (8)$$

Otherwise, the scattered lights from different atoms will cancel each other. This relationship corresponds to conservation of energy and momentum in quantum picture, with the + sign corresponding to creation of a phonon and is called stokes process, while the - sign corresponds to anti-stokes process. When the above conditions are satisfied, the electric field equation of Raman scattered light for bulk crystals can be simplified into:

$$\vec{E}_s\propto\sum_{ab}A_a\hat{e}_b\left(\hat{e}_b\cdot\overleftrightarrow{R}\cdot\hat{e}_a\right)A_{(\vec{q},j)}e^{i(\omega_i\pm\omega(\vec{q},j))t} \qquad (9)$$

Where $\overleftrightarrow{R}\equiv\sum_{fkd}\frac{g_{kd}(\vec{q},j)}{\sqrt{m_k}}\frac{\partial\vec{\overleftrightarrow{x}}_k}{\partial r_{kd}}$ is the traditional Raman tensor. Putting the above equation back into the previous dipole radiation intensity equation will give the Raman scattering intensity. The ratio between the Raman intensities of Stokes and anti-Stokes process will be:

$$\frac{I_{Stokes}}{I_{anti-Stokes}}=\frac{{\omega_{Stokes}}^4}{{\omega_{anti-Stokes}}^4} \qquad (10)$$

Where $\omega$ is the frequency of the Raman scattered light. It seems from the above equation that the stokes process and anti-stokes process will have similar intensity because the phonon frequencies are usually much smaller than the frequency of incident light. Unfortunately, this is not true. This is a limitation of classical Raman scattering theory because it cannot predict the Raman intensity of stokes process and anti-stokes process correctly [3,4]. The correct Raman intensity from quantum theory should be [3,4]:

$$\mathrm{I} \propto \frac{\omega_s^4}{\omega(\vec{q},j)} * \begin{cases} (\bar{n}+1) & (stokes) \\ \bar{n} & (antistokes) \end{cases} \qquad (11)$$

where $\bar{n}$ is the mean occupation number of the particular phonon and it follows Bose–Einstein distribution. Generally, at room temperature $\bar{n}$ will be very close to zero so experiments usually measure stokes process only. Unfortunately, this paper does not have the ability to derive the birefringent version of quantum Raman scattering theory so it does not know how birefringence would affect the stokes and anti-stokes Raman scattering intensities. But it may be safe to assume that the Raman intensities will remain similar as the isotropic cases.

## Discussion

It can be seen from the above derivations that there is no simple rule regarding Raman intensities. It is a complicated interplay between refractive index, phonon wave vector and crystal size. But the rule of thumb is that the Raman scattered light will split into four different polarized parts with each one being incoherent to each other because they come from scattering with different phonons. And the intensities of each of these incoherent parts can be calculated using the traditional way of Raman tensors. But the frequency differences between them may not be discernible because the first order derivative of $\omega(\vec{q},j)$ at Gamma point is 0 due to $\omega(\vec{q},j)$ being an even function. Normally the wavelength of incident light is much larger than the size of the primitive cell of crystal, thus as a result the Raman active phonon wave vector will be very close to the center of Brillouin zone or the Gamma point.

The interference effect is contained in equation 7 where it sums up individual Raman contribution from all atoms. When the sample size is infinite, the phonons must satisfy a phase matching condition as specified in equation 8 and when this condition is satisfied there will not be any interference effect because all individual contributions of Raman scattering from each atom will be perfectly in phase with each other. However, when the sample size is limited, then the phase matching condition can no longer be exactly satisfied since the phonon dispersion curve will no longer be continuous, and interference will occur between Raman scatterings from each atom. This is something that can be inferred from the reference derivations [3,4] but it was generally ignored because layered materials were not a thing back then. For layered materials the phonon dispersion curves alone the directions parallel to the surface of the material may still be continuous. This probably will mitigate the interference effect so that thickness will not affect too much of the Raman profile.

Angle resolved polarized Raman spectroscopy is where birefringence can significantly affect the spectral profiles. In the common backscattering configuration, if the Raman tensor $\overleftrightarrow{R}$ is of the form $\begin{matrix} 1 & 0 & 0 \\ 0 & 1 & 0 \\ 0 & 0 & * \end{matrix}$, then the angle resolved polarized Raman intensity profile for a bulk isotropic material with the polarization direction of scattered light $\hat{e}_s$ in the XY plane and parallel to that of incident light $\hat{e}_i$ will be $\mathrm{I}_{\parallel}(\theta) \propto \left|\hat{e}_s \cdot \overleftrightarrow{R} \cdot \hat{e}_i\right|^2 = 1$. But for bulk birefringent crystals in backscattering, according to the theory of this paper of incoherent assumption, it will become: $\mathrm{I}_{\parallel}(\theta) \propto \left|\cos\theta\, \hat{e}_o \cdot \overleftrightarrow{R} \cdot \hat{e}_o \cos\theta\right|^2 + \left|\sin\theta\, \hat{e}_e \cdot \overleftrightarrow{R} \cdot \hat{e}_e \sin\theta\right|^2 + \left|\sin\theta\, \hat{e}_e \cdot \overleftrightarrow{R} \cdot \hat{e}_o \cos\theta\right|^2 + \left|\cos\theta\, \hat{e}_o \cdot \overleftrightarrow{R} \cdot \hat{e}_e \sin\theta\right|^2 = sin^4\theta + cos^4\theta$, where $\hat{e}_o$ represents the polarization direction of the o ray and $\theta$ is the angle between the incident polarization direction and the o ray polarization direction. However, the caveat is that if one assumes the different polarized parts to be out of phase instead rather than to be incoherent as the literatures did [5-10], then the same profile equation can still be obtained. Thus, it can be said that this theory does not provide immediate benefits over literature in terms of being more agreeable with experiment results but rather it just gives a different explanation for the same phenomenon. On the other hand, this can also be taken in another way that this paper at least has not contradicted existing experimental results. This is because the existing theories in literature certainly have agreed with the existing experiments that they addressed. If this paper predicts a different polarized Raman spectral shape than them, then it would mean that this paper may be contradicting with those experimental results. Nevertheless, there must be some discernible different experimental predictions between the theory of this paper and the literature theories, otherwise this paper would become useless. There indeed are some differences predicted from this paper. From the literature theories which assume that the Raman scattering lights at different polarization angles are coherent with each other, it can be predicted that the polarized Raman profile shape must be dependent on the thickness of the sample if the incident light can penetrate the full depth of the sample. However, according to the theory of this paper, if the Raman scattering at different polarization angles are incoherent with each other, then the polarized Raman profile will not depend on sample thickness if the interference effect is ignored. Thus, this phenomenon can potentially be used to validate or invalidate this paper. However, if the material is only weakly birefringent, then the Raman scattering at different polarization angles will only be partially incoherent. This is because when the birefringence is weak, the lattice vibration modes responsible for one Raman polarization direction will also be able to generate Raman emission at another polarization direction albeit at reduced intensities. As a result, for weak birefringent materials both the literature theories and this paper will predict thickness dependence of polarized Raman profile to some weaker degrees. Another problem is that if the sample size is limited, then there will be interference effects which are dependent on sample size. Thus, it will only be with sufficiently strong birefringent materials and suitable sizes that one can find qualitatively different predictions from the literature theories and the theory of this paper. Nevertheless, given that the parameters of the theory of this paper have very different physical meanings than the literature, it may be more suitable for ab initio calculations of polarized Raman profiles or at the very least help provide a new perspective on understanding the physics behind Raman scattering in birefringent materials.

Another prediction by the theory of this paper is that the scattered Raman light can have different Raman shifts at different polarization directions for birefringent materials. The is because the Raman emissions at different polarization angles arise from interactions with different lattice vibration modes separately. Thus, they are independent from each other and free to have different frequencies. However, the difference in Raman shifts between different polarization angles if existed must be very small because the first order derivative of phonon dispersion curve (phonon frequency versus phonon wave vector) at Gamma point is zero due to it being an even function. Although there may be other factors that would increase the Raman shift differences between polarization angles. For example, the forces on atoms in lattice vibration are not exactly harmonic but contain some higher order non-harmonic components. And for samples of limited thickness such as the layered materials the phonon mode distribution in phonon dispersion curves is not continuous in the direction perpendicular to sample plane. Nevertheless, this paper did happen to have found an experimental result from literature of polarized Raman spectroscopy profile of black phosphorus that supports this prediction [12]. One can see from the figure 3 and 4 in this literature paper as reproduced in figure 1 below that the Raman shifts of zigzag and armchair polarization direction are consistently different across a range of temperatures although this difference is very small compared to the full-width-at-half-maxima of the Raman spectral lines. Actually, the purpose of these literature figures was to show the temperature dependence of Raman frequencies which was due to anharmonicity effect. The half-maxima width is also attributable to this effect. This phenomenon of different Raman peak frequences at different polarization directions for birefringent materials was never addressed by literature and this paper posits that this phenomenon can be useful for identifying the orientation of crystals. It should be noted that there could be other explanations for this polarization dependence of Raman shifts. For example, the anisotropic sample may heat up at different rates under different polarization angles [13]. The slope of the Raman shift to temperature curves in figure 1 means that the temperature difference under different polarization angles needs to be tens of degrees to be responsible for the polarization dependence of Raman shifts. But according to the reference paper measures were taken to avoid sample heating: "*For the polarization dependence measurement, the sample was placed on a rotation stage and was rotated during the measurement. The laser power was kept at below 1 mW on the sample in order to avoid thermal heating caused by the laser beam. A 100× magnification objective was used to focus the laser beam into about 1 μm*." Thus, further research is probably needed to find out for sure whether anisotropic thermal absorption can be a reason for the Raman shift difference between different polarization angles.

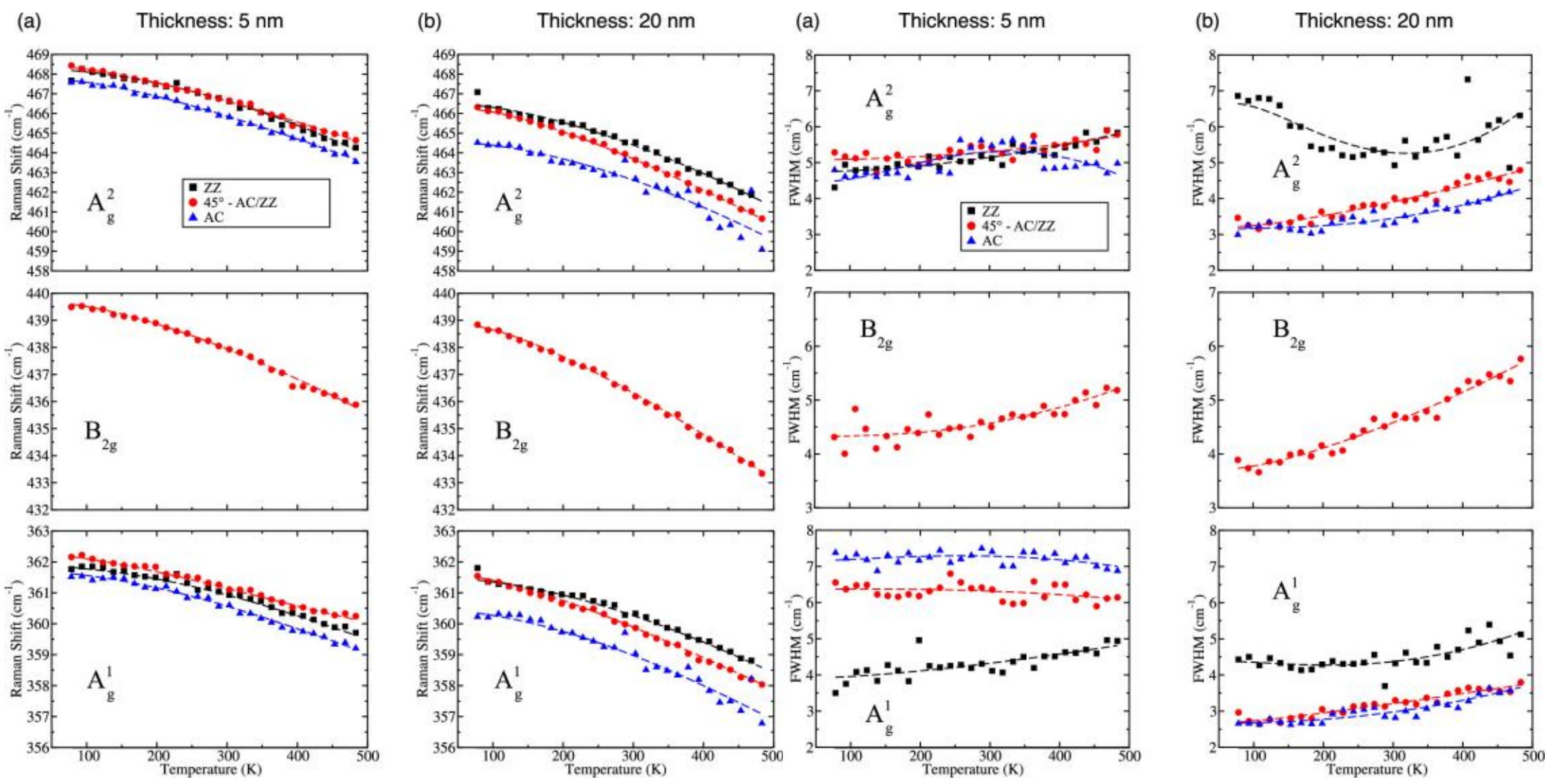

*Figure 1 Temperature/polarization dependence of $A_g^1$, $B_{2g}$, and $A_g^2$ Raman shift frequencies and full-width-at-half-maxima of few-layer of black phosphorus for two thicknesses: (a) 5 nm and (b) 20 nm at incident polarization angles of zigzag (ZZ), armchair (AC), and 45° between AC and ZZ directions respectively.*

## Conclusion

This paper has successfully developed a classical Raman scattering theory from first principle that can quantitively assess birefringence and interference effects in anisotropic materials. The theory gives different predictions and interpretations to the birefringence phenomena in Raman spectroscopy than the literature that this study can find. It can also be used for ab initio calculations. It is hoped this theory can help shed new light on the birefringence effect in Raman scattering for future theoretical and experimental studies.

## Declaration

No new data is generated in this paper.

There are no known conflicts of interest.